\documentclass[twocolumn]{aastex631}
\usepackage{amsmath}
\usepackage{color}
\usepackage{natbib}
\usepackage{amssymb}
\usepackage{amsfonts}
\usepackage{stackengine}
\usepackage{changepage}
\usepackage{subfigure}
\usepackage{graphicx}
\usepackage{rotating}
\usepackage{threeparttable}
\usepackage{xspace}
\usepackage{soul}
\newcommand{\beq}{\begin{equation}}
\newcommand{\eeq}{\end{equation}}

\def\ba{\begin{eqnarray}}
\def\ea{\end{eqnarray}}

\shorttitle{ }

\shortauthors{Peng et al.}
\begin{document}

\title{The Head-on Collision of a Neutron Star with a White Dwarf}

\correspondingauthor{Zong-kai Peng}
\email{zkpeng@bnu.edu.cn}

\correspondingauthor{He Gao}
\email{gaohe@bnu.edu.cn}


\author[0000-0003-1904-0574]{Zong-kai Peng}
\affiliation{Institute for Frontier in Astronomy and Astrophysics, Beijing Normal University, Beijing 102206, China}
\affiliation{Department of Astronomy, Beijing Normal University, Beijing 100875, China}

\author[0000-0002-3100-6558]{He Gao}
\affiliation{Institute for Frontier in Astronomy and Astrophysics, Beijing Normal University, Beijing 102206, China}
\affiliation{Department of Astronomy, Beijing Normal University, Beijing 100875, China}

\author{Xian-Fei Zhang}
\affiliation{Institute for Frontier in Astronomy and Astrophysics, Beijing Normal University, Beijing 102206, China}
\affiliation{Department of Astronomy, Beijing Normal University, Beijing 100875, China}

\begin{abstract}

We have computed the physical processes involved in a head-on collision between a neutron star (NS) and a white dwarf (WD). The outcomes of such collisions vary depending on the mass and type of the WD. We have separately examined the dynamical processes for collisions between NSs and helium WDs (He-WDs), carbon-oxygen WDs (CO-WDs), and oxygen-neon WDs (ONe-WDs). We aim to investigate whether the collision can trigger a thermonuclear explosion of the WD, and if not, whether the NS can remain bound within the WD to form a Thorne–Żytkow-like object (TZlO). For a thermonuclear explosion to occur, at least two conditions must be satisfied: (i) the collision-induced temperature must reach the ignition threshold of the relevant nuclear reactions, and (ii) the burning material must remain in a degenerate state. For different types of WDs, there exist parameter ranges where both conditions are fulfilled, implying that NS–WD collisions can indeed induce thermonuclear explosions, leading to sub-Chandrasekhar Type Ia supernovae or other exotic optical transients powered by thermonuclear explosion. On the other hand, the formation of a TZlO requires that the WD material exerts sufficient drag on the NS to prevent its escape, while the interaction must not trigger a thermonuclear explosion of the WD. Our results indicate that such conditions can be realized in the case of low-mass CO-WDs and low-mass ONe-WDs, provided that the viscous coefficient is sufficiently large.

\end{abstract}

\section{Introduction}

In recent years, binary neutron star-white dwarf (NS-WD) systems have increasingly attracted the attention of researchers. For instance, \citet{Yang2022} proposed that the peculiar gamma‐ray burst GRB 211211A may originate from the merger of an NS with a WD. \citet{Zhong2023} also interpreted the burst in terms of an NS–WD merger, albeit invoking a different explosion mechanism.

Before these studies,  \citet{Metzger2012}, \citet{Margalit2016, Margalit2017}, \citet{Zenati2020}, and \citet{Bobrick2017, Bobrick2022} did numerical simulations to investigate nucleosynthesis and mass transfer during NS–WD mergers. Their results indicate that the merger process can synthesize small amounts of nickel (Ni), which may lead to relatively faint optical transients, such as some SNe Iax \citep{Bobrick2022} and Ca‑rich SNe \citep{Margalit2016}.

In addition, \citet{Pas2011b} used general relativistic hydrodynamic simulations to study the NS–WD merger process, finding that the NS ultimately falls into the WD, resulting in the formation of a Thorne–Żytkow-like object (TZlO). Based on these simulation results, in our previous work, we discussed in detail the formation of the TZlO and its subsequent evolution, successfully accounting for the observational features of both the “long–short burst” GRB 211211A and the “short–long burst” GRB 200826A \citep{Peng2024}.

On the other hand, \citet{Pas2011a} computed a head-on collision between NS and WD, proposing a more direct route to forming a TZlO. However, to simplify the calculations, they employed a pseudo-WD in place of a realistic WD and did not consider whether the collision could trigger an explosion of the WD. Motivated by this limitation, we examined the collision process between NS and WD at realistic scales. By solving the equations of motion for collisions involving NS and various types and masses of WD, we found that the NS does not always remain within the WD, in many cases, it transpierces the WD. Further calculations indicate that the high temperatures generated in the interaction region as the NS passes through are sufficient to ignite the nuclear burning of the WD material, thereby triggering a thermonuclear explosion. We thus propose that NS–WD collisions may represent an extremely rare channel for producing SNe Ia. This mechanism is analogous to the WD–WD head-on collision model for SNe Ia \citep{Rosswog2009, Raskin2009}, however, given the much lower abundance of NS compared to WD and the smaller collision cross-section, the probability of NS–WD collisions is naturally lower than that of WD–WD collisions. There are also cases in which, following the collision, the NS remains embedded within the WD without triggering a thermonuclear explosion. This indicates that head-on NS–WD collisions indeed represent a viable formation channel for TZlOs.

The lack of discussion surrounding such systems is primarily due to the extremely low event rate. Although we know that the event rate of NS-WD collisions must be very low, estimating this rate is not straightforward. As a collision event similar to NS-WD collisions, the calculation of the event rate for WD-WD collisions still carries significant uncertainty, however, it provides us with a rough benchmark for our expectations regarding the NS-WD collision rate. \citet{Katz2012} argued that the collision rate of double WDs could be high enough to be comparable to the observed rate of SNe Ia, whereas \citet{Toonen2018} suggested that the probability of WD-WD collisions in triple systems is much lower than the SNe Ia event rate. \citet{Dong2015, Vallely2020, Tucker2025} identified several candidate sources in a Type Ia supernova sample that may have originated from WD–WD collisions, which in turn raises our expectations for NS–WD collisions. Although the probability of NS–WD collisions is much lower than that of WD–WD collisions, we consider that the occurrence rate should not be so low as to be entirely negligible. Therefore, we believe it is worthwhile to investigate the physical effects of NS–WD collisions.

This paper is organized as follows. We calculate the kinematic equation of the NS-WD collision process and the temperature curve of the interaction region in \S 2. In \S 3, we discuss the NS collide with the different types of WD. A brief summary and discussion follow in \S 4.

\section{The Physical Model}
In the NS-WD collision process, we are interested in two specific scenarios: one is inspired by \citet{Pas2011a}, which considers the case where the relative velocity between the NS and WD material during the collision is extremely high. As the NS traverses the WD, it experiences substantial drag force. We aim to investigate whether this drag effect can sufficiently decelerate the NS, ultimately allowing it to remain within the WD, potentially forming a TZlO system. The other scenario is whether the collision and subsequent traversal can generate sufficiently high-temperature hotspots near the interaction region to trigger a thermonuclear explosion in the WD. 

\subsection{The NS-WD collision process}

Consider a WD of mass $M_{\rm WD}$ and an NS of mass $M_{\rm NS}$ approaching each other from a considerable distance. When the WD reaches the tidal disruption radius ($a_{0}$) of the NS, the gravitational influence of the NS becomes significant enough to cause material from the WD to flow toward the NS. At this point, the relative velocity between the two stars is given by $v = [2G(M_{\rm WD}+M_{\rm NS})/a_{0}]^{1/2}$. Neglect the deformation of the WD, the NS collides with the WD at the distance between the two reaches $R_{\rm WD}+R_{\rm NS}$. Then, the relative velocity at the moment of collision is approximated as
\begin{eqnarray} 
v_{\rm r} = [2G(M_{\rm WD}+M_{\rm NS})/(R_{\rm WD}+R_{\rm NS})]^{1/2}. 
\end{eqnarray}
The NS plunges into the WD would make a strong shock wave, at the same time, the WD material creates a strong drag force on the NS. For the case of the NS traject inside the WD, the material of WD could be treated as fluid. The drag force worked on the NS could be expressed as $F_{\rm d}=\frac{1}{2}C_{\rm d} \rho_{\rm WD}v^{2}A$, where $C_{\rm d}$ is the drag coefficient, $\rho_{\rm WD}$ is the density of WD, the $v$ is the relative velocity of the NS and the WD material, the $A$ is the interaction cross-section. $C_{\rm d}$ is a dimensionless parameter related to the shape of the moving object, the physical character of the fluid, the Mach number, etc. It is challenging to estimate the drag coefficient of an NS traveling through the electron degeneracy pressure supported WD material. As the NS traverses through the interior of the WD, it generates strong shocks in the direction of its motion, resulting in an interaction cross-section that is larger than the geometric cross-section of the NS. We assume $A = \pi (n R_{\rm NS})^{2} = n^{2}A_{\rm NS}$, where $A_{\rm NS}$ is the cross-section of the NS. Considering that the relative velocity of the NS inside the WD may exceed the local sound speed, the larger the Mach number of this supersonic motion, the stronger the resulting shock effects, leading to a more significant deceleration of the NS. The Mach number could be estimated to be $\mathcal{M} = v_{\rm r}/c_{\rm s,WD}$, where 
\begin{eqnarray} 
c_{\rm s,WD} = \left(\frac{\partial \gamma P_{\rm WD}}{\partial \rho_{\rm WD}}\right)_{s}\sim \frac{\gamma P_{\rm WD}}{\rho_{\rm WD}}
\end{eqnarray}
is the sound speed of the WD. The $\gamma$ is the polytropic exponent. The $P_{\rm WD}$ is the internal pressure of the WD, which can be expressed as
\begin{eqnarray} 
P_{\rm WD}\sim \frac{GM_{\rm WD}^{2}}{R_{\rm WD}^{4}}.
\end{eqnarray}
Then we have
\begin{eqnarray} 
\mathcal{M}=\frac{v_{\rm r}}{c_{\rm s,WD}}\sim (1+q)^{1/2},
\end{eqnarray}
the $q = \frac{M_{\rm NS}}{M_{\rm WD}}$. We assume that the drag effect scales positively with the Mach number, Then $F_{\rm d} = \frac{1}{2}n^{2}(1+q)^{1/2}\tilde{C}_{\rm d} \rho_{\rm WD} v^{2}A_{\rm NS}$\footnote{In fact, the Mach number evolves during the passage of the NS through the WD. For massive WDs, the drag effect exerted on the NS is significant, leading to a relatively smaller Mach number. In contrast, collisions involving less massive WDs correspond to larger Mach numbers, while their deceleration effect on the NS is weaker, resulting in only minor variations of the Mach number throughout the traversal.}. We introduce an effective drag coefficient $C_{\rm n} = n^{2} \tilde{C}_{\rm d}$ as a free parameter on account these two parameters are degenerate. The drag force could be expressed as
\begin{eqnarray} 
F_{\rm d} = \frac{1}{2}(1+q)^{1/2}C_{\rm n} \rho_{\rm WD} v^{2}A_{\rm NS}.
\end{eqnarray}

Considering the gravitational attraction and drag force acting between the NS and WD post-collision, we set up the equations of motion in the center-of-mass coordinate, using the initial conditions at the collision point

\begin{eqnarray}
\begin{cases}
\begin{aligned}
\frac{d^{2}\vec{\textbf{r}}_{1}}{dt^{2}} = &-\frac{3}{8}(1+q)^{1/2}\frac{C_{\rm n}R_{\rm NS}^{2}}{R_{\rm WD}^{3}}\frac{(\vec{\textbf{v}}_{1} - \vec{\textbf{v}}_{2})^{3}}{|\vec{\textbf{v}}_{1} - \vec{\textbf{v}}_{2}|}  \\
&+ \frac{GM_{\rm NS}}{R_{\rm WD}^{3}}(\vec{\textbf{r}}_{2}-\vec{\textbf{r}}_{1}) \\[1ex]
\frac{d^{2}\vec{\textbf{r}}_{2}}{dt^{2}} = &\frac{3}{8}(1+q)^{1/2}\frac{C_{\rm n}M_{\rm WD}}{M_{\rm NS}}\frac{R_{\rm NS}^{2}}{R_{\rm WD}^{3}}\frac{(\vec{\textbf{v}}_{1} - \vec{\textbf{v}}_{2})^{3}}{|\vec{\textbf{v}}_{1} - \vec{\textbf{v}}_{2}|}  \\
&- \frac{GM_{\rm WD}}{R_{\rm WD}^{3}}(\vec{\textbf{r}}_{2}-\vec{\textbf{r}}_{1}),
\end{aligned}
\end{cases}
\end{eqnarray}
where $\vec{\textbf{r}}_{1}$ and $\vec{\textbf{r}}_{2}$ are the position vectors of the centers of the WD and NS, respectively. 
$\vec{\textbf{v}}_{1} = \frac{d\vec{\textbf{r}}_{1}}{dt}$ and $\vec{\textbf{v}}_{2} = \frac{d\vec{\textbf{r}}_{2}}{dt}$ are the velocity vectors of the WD and NS, respectively. Once the NS exits the interior of the WD, the interaction between the two objects reduces to pure gravity, and the governing equations simplify accordingly:

\begin{eqnarray} 
\begin{cases}
\begin{aligned}
\frac{d^{2}\vec{\textbf{r}}_{1}}{dt^{2}} &= GM_{\rm NS}\frac{\vec{\textbf{r}}_{2}-\vec{\textbf{r}}_{1}}{|\vec{\textbf{r}}_{2}-\vec{\textbf{r}}_{1}|^{3}} \\[1ex]
\frac{d^{2}\vec{\textbf{r}}_{2}}{dt^{2}} &= -GM_{\rm WD}\frac{\vec{\textbf{r}}_{2}-\vec{\textbf{r}}_{1}}{|\vec{\textbf{r}}_{2}-\vec{\textbf{r}}_{1}|^{3}}.
\end{aligned}
\end{cases}
\end{eqnarray}
The initial conditions can be expressed as:
\begin{eqnarray} 
\begin{cases}
\begin{aligned}
&r_{1}(t=0) = -\frac{M_{\rm NS}}{M_{\rm NS} + M_{\rm WD}}R_{\rm WD}\\
&r_{2}(t=0) = \frac{M_{\rm WD}}{M_{\rm NS} + M_{\rm WD}}R_{\rm WD}\\
&\frac{dr_{1}}{dt}|_{t=0} = \frac{M_{\rm NS}}{M_{\rm NS} + M_{\rm WD}}v_{r}\\
&\frac{dr_{2}}{dt}|_{t=0} = -\frac{M_{\rm WD}}{M_{\rm NS} + M_{\rm WD}}v_{r}.
\end{aligned}
\end{cases}
\end{eqnarray}
Fixing the neutron star mass at 1.4$M_{\odot}$, the primary parameters influencing the equations of motion are $C_{\rm n}$ and the mass of the WD.

\subsection{The model testing}

\begin{figure*}[ht]
    \includegraphics[width=\textwidth]{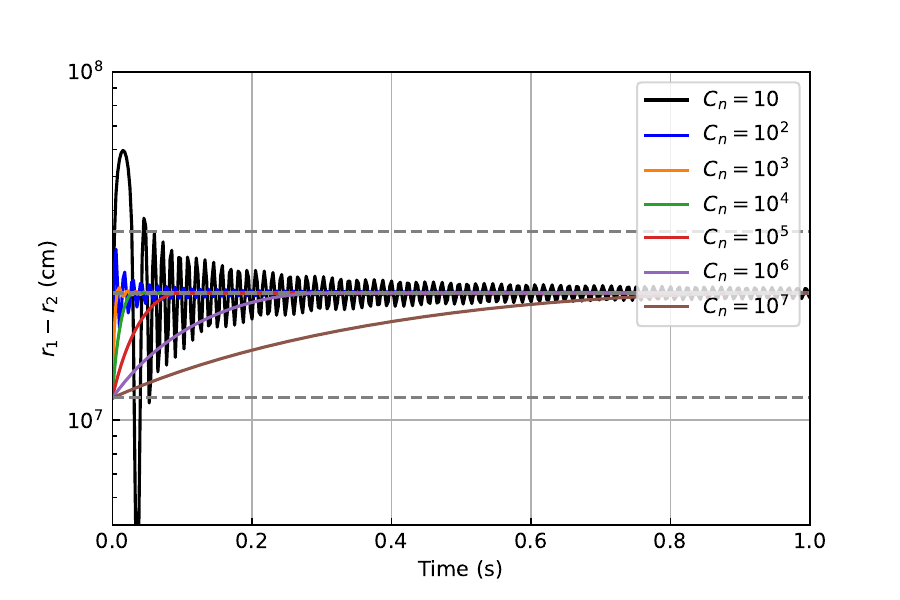}
    \caption{Evolution of the relative distance between the pseudo–WD and the NS after the collision, for a WD mass of $1.0~M_{\odot}$. The solid lines represent the relative distance for seven different values of $C_{\rm n}$ , while the dashed line denotes the sum of the WD and NS radii. Due to the energy loss during the collision, all results consistently show that the NS ultimately sinks toward the center of the WD.}
    \label{fig:Rela}
\end{figure*}

Due to the simplifications adopted in our model, we test its validity by following the approach of \citet{Pas2011a}, where the size of the WD is artificially compressed. Specifically, we take the mass of the WD to be $1.0M_{\odot}$ and reduce its radius to ten times the NS radius, consistent with the parameters of model A2 in \citet{Pas2011a}. Under these conditions, the evolution of the relative position between the NS and the center of the pseudo-WD is shown in Figure \ref{fig:Rela}. As evident, for $C_{\rm n}>10^2$, the energy loss from the collision is significant enough that the NS cannot penetrate the WD. Instead, it undergoes oscillations near the WD's core and eventually settles at the center. For $10^2<C_{\rm n}<10^{5}$, the NS remains near the core for a characteristic timescale of approximately 0.1~s, which is consistent with the results shown in Figure 4 of \citet{Pas2011a}. In cases with $C_{\rm n}>10^{5}$, the deceleration due to the collision becomes more severe, and the NS takes longer to reach the WD’s core. For example, at $C_{\rm n}=10^{6}$, the NS takes roughly 0.3~s to reach the center, while for $C_{\rm n}=10^{7}$, this time increases to about 0.8~s. These results indicate that within the parameter range $10^2<C_{\rm n}<10^{5}$, our model yields behavior consistent with that of \citet{Pas2011a}.

\subsection{The NS collide with the real scale WD}

\begin{figure*}[ht]
	\centering
	\vspace{-0.15in}
	\begin{minipage}{1\linewidth}	
		\subfigure{
			\label{fig:1}
			\includegraphics[width=0.49\linewidth,height=2.5in]{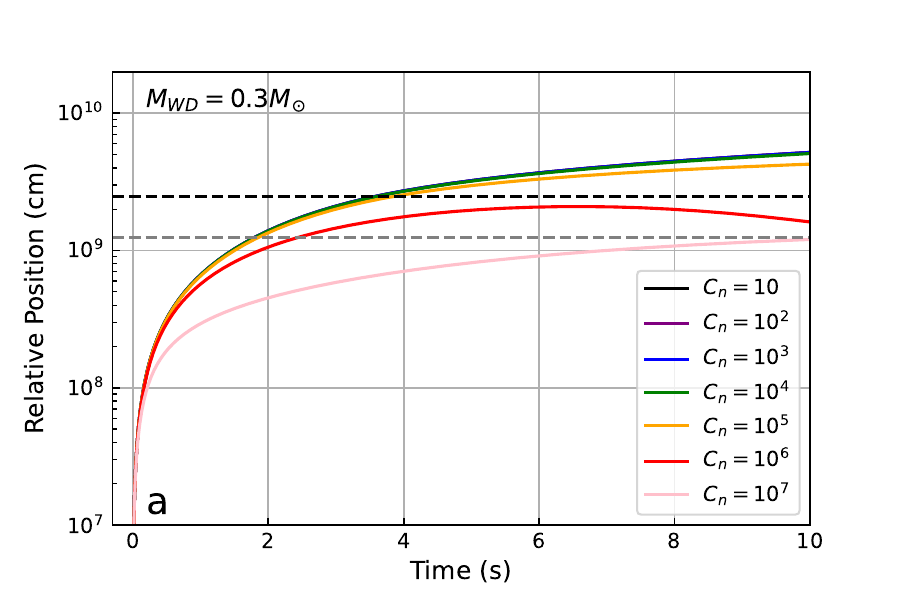}	
		}\noindent
		\subfigure{
			\label{fig:2}
			\includegraphics[width=0.49\linewidth,height=2.5in]{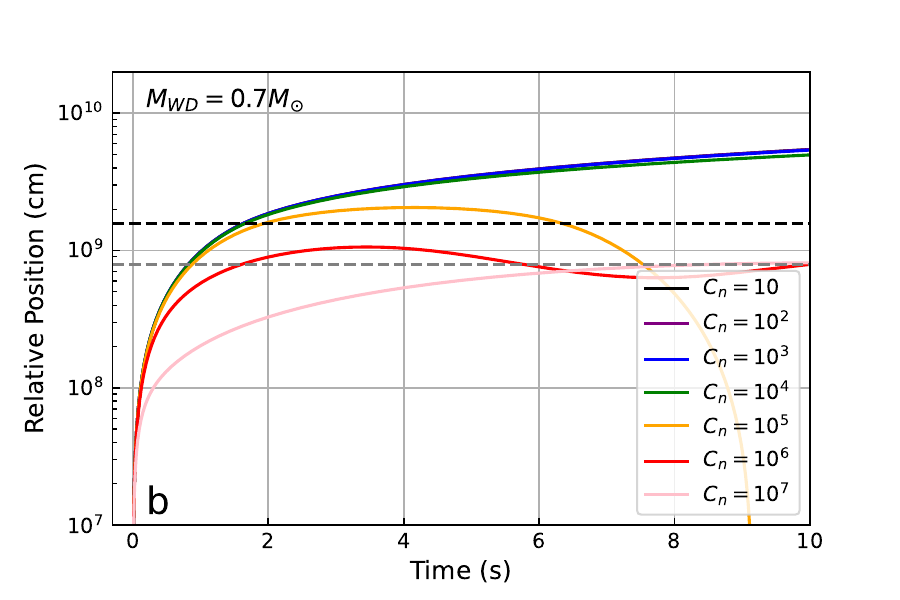}
		}
	\end{minipage}
	\vskip -0.8cm 
	\begin{minipage}{1\linewidth }
		\subfigure{
			\label{fig:3}
			\includegraphics[width=0.49\linewidth,height=2.5in]{{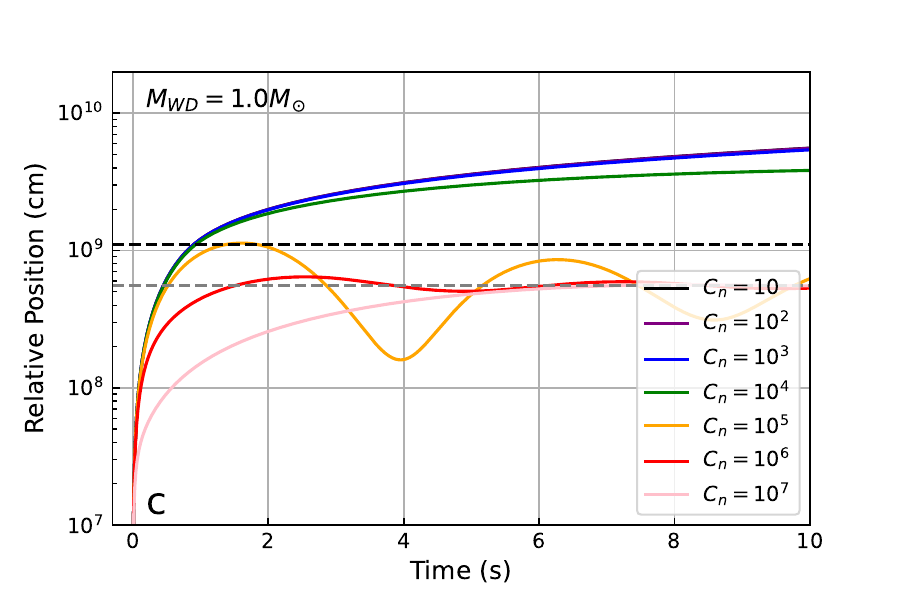}
			}
		}\noindent
		\subfigure{
			\label{fig:4}
			\includegraphics[width=0.49\linewidth,height=2.5in]{{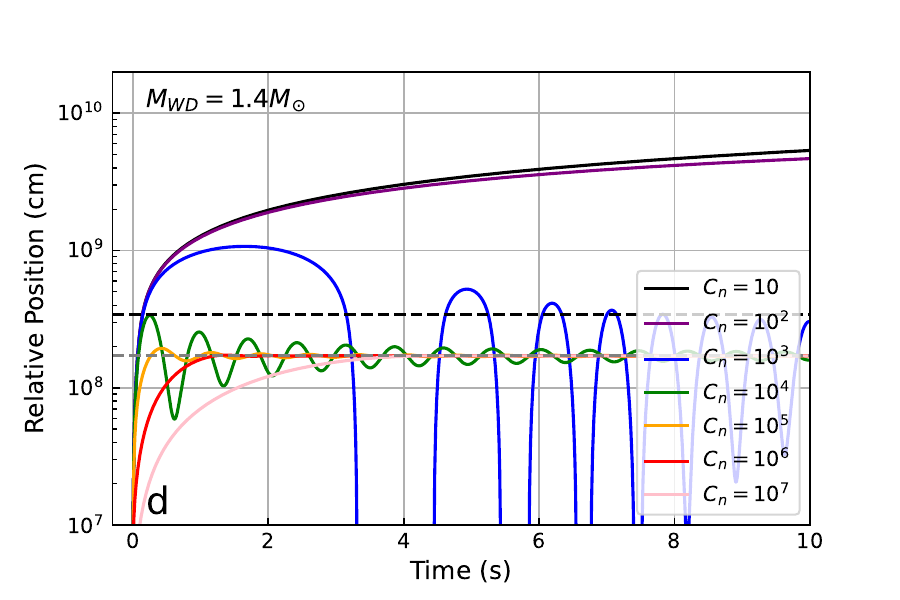}
			}
		}
	\end{minipage}
	\vspace{-0.18in}	
	\caption{The relative displacement of NS with the center of WD for different parameters. For the purpose of visualization, a constant offset was applied to the relative distance such that the initial separation is normalized to zero. The light gray dashed line denotes the coincidence of the NS and WD centers of mass, while the black dashed line indicates the condition where the separation between the two centers of mass exceeds the sum of their radii, corresponding to the NS emerging from the WD.}
	\vspace{0.1in}		
	\label{fig:Disp}
\end{figure*}

\begin{figure*}
    \centering
    \includegraphics[width=\textwidth]{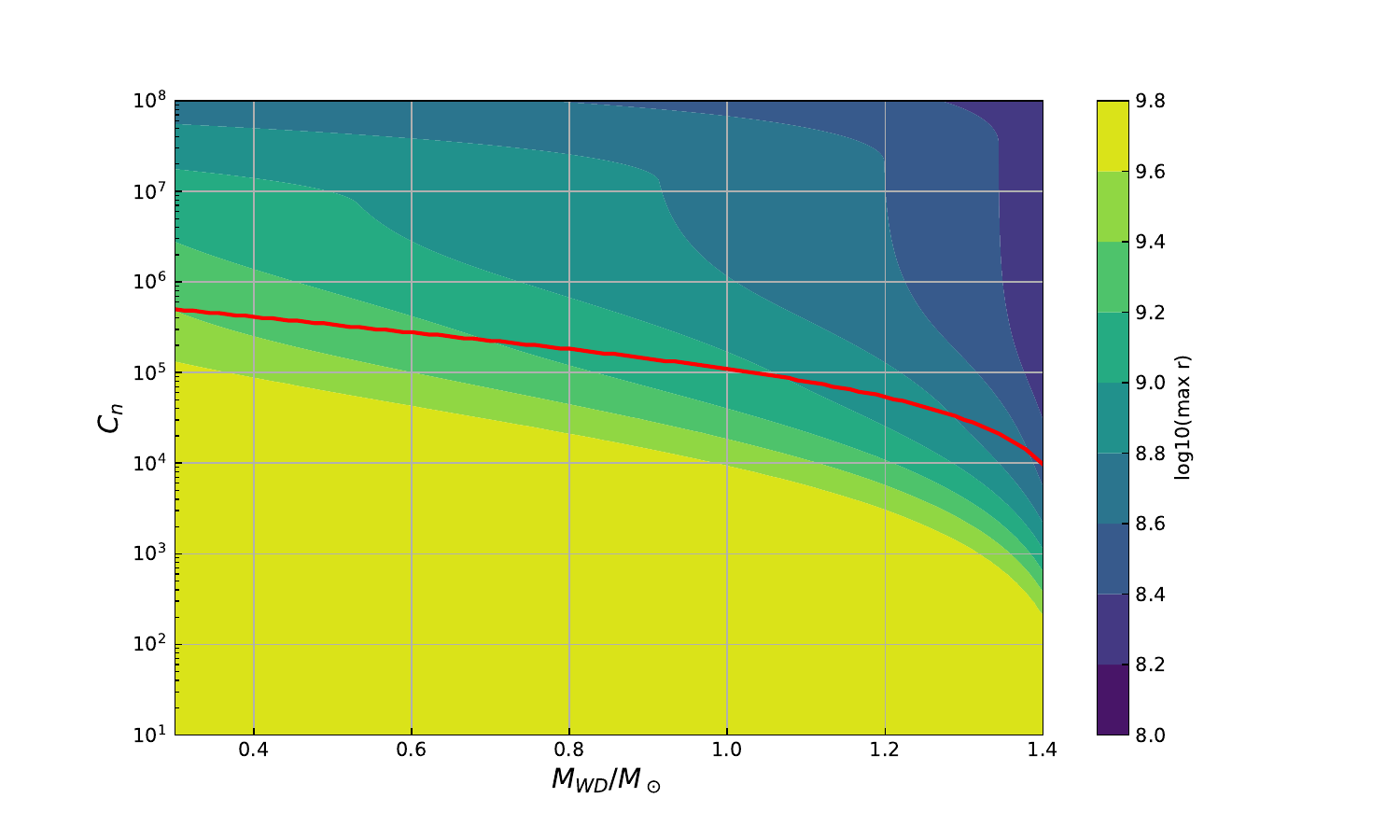}
    \caption{Taking the collision between the NS and WD as the initial moment, we measure the maximum separation between them during their relative motion over the subsequent 10 seconds. The red line indicates that the NS was just able to pierce through the WD.}
    \label{fig:coll}
\end{figure*}

Figure \ref{fig:Disp} shows the displacement curves for different values of $C_{\rm n}$ and varying WD masses. It can be seen that smaller values of WD mass and $C_{\rm n}$ correspond to a reduced deceleration effect on the NS. To examine the parameter conditions under which the NS can be sufficiently decelerated and thus prevented from escaping the WD, we evaluate the maximum relative distance between the two objects over 10 seconds across various parameters. The results are shown in Figure \ref{fig:coll}, where the red line represents the parameter distribution curve for which the maximum distance within 10 seconds equals the WD's diameter. It can be seen that whether the NS can stay inside the WD depends on $C_{\rm n}$ and the mass of the WD.

It is critical to emphasize that our calculations are based on the dynamics under idealized conditions. Under certain parameter regimes, the NS may exit the WD and then be drawn back by gravity to re-enter it, while in other cases the NS may be sufficiently decelerated within the WD to ultimately oscillate near the WD core. In realistic scenarios, however, an NS-WD collision would result in a complete alteration of the WD's physical state. For example, the shockwave produced by the collision may entirely disrupt the WD. Although \citet{Pas2011a} did not report such an outcome, \citet{Van2024} performed numerical simulations of WD–main-sequence star collisions that demonstrated complete stellar disruption, suggesting that the post-collision fate of the NS and WD warrants careful consideration. Another possibility, as will be discussed in Section 3, is that the high temperatures generated during the collision could ignite the WD, leading to a thermonuclear explosion. The explosion’s shockwave might also influence the NS's trajectory. Moreover, since the timescales for the WD explosion and the NS traversing the WD are comparable, the resulting supernova creates an explosively reshaped mass distribution that could further affect the NS's motion. In addition, we ignore the effect of angular momentum on the motion of the system. These issues lie beyond the scope of the present work and should be addressed in future numerical simulations for a more detailed analysis.

\subsection{The temperature of the hotspot}
The relative velocity at the moment of collision between the NS and the WD is $\sim 10^{9}\rm~cm~s^{-1}$, and such an intense collision can easily generate hotspot\footnote{The hotspot discussed here is not a fixed point. Since it arises from the interaction between the NS and the white dwarf material, it is produced in the region where these two bodies interact and continuously evolves with the motion of the NS.} at the interaction region. The WD is supported by electron degeneracy pressure, and due to the long mean free path of electrons, it has a high thermal conductivity. Assuming that the electrons within the WD move at relativistic speeds, we set the thermal conduction rate at the speed of light to study the local temperature evolution as the NS traverses through the WD.

At the time $t_{i}$, the average temperature near the NS can be described as the overall effect on the internal energy of that region due to the energy dissipated from the initial moment up to $t_{i}$. Given that the thermal conduction speed greatly exceeds the velocity of the NS, the temperature at the interaction region at $t_{i}$ is primarily influenced by the energy dissipated during the interval from $t_{i-1}$ to $t_{i}$. The total energy loss from $t_{i-1}$ to $t_{i}$ is $\Delta E_{i} = \Delta E_{p,i} + \Delta E_{k,i}$, where $\Delta E_{p,i}$ is the potential energy loss, $\Delta E_{k,i}$ is the kinetic energy loss. Considering that a large portion of this lost energy is converted into the kinetic energy of white dwarf material clumps, with only a fraction converted into internal energy, we assume a conversion efficiency $\eta$. Thus, the contribution to the temperature at that location due to energy converted to internal energy from $t_{i-1}$ to $t_{i}$ can be expressed as $\Delta T_{i,1} = (\frac{\eta \Delta E_{i}}{a \Delta V})^{1/4}$, where $a$ is the radiation constant, $\Delta V = \frac{4}{3}\pi (c \Delta t)^{3}$ is the volume element affected by thermal conduction within the time $\Delta t$\footnote{Consider that the scale of thermal conduction within $\Delta t$ should be compared with the scale of the NS, we set $\Delta t = 0.1~\rm ms$.}. The actual temperature in this region should also include the cumulative effect from the initial moment up to $t_{i-1}$. The contribution to the temperature from the energy loss converted to internal energy during the interval $t_{i-2}$ to $t_{i-1}$ is $\Delta T_{i,2} = (\frac{\eta \Delta E_{i-1}}{2^{3}a \Delta V})^{1/4}$, indicating that the earlier energy losses have progressively smaller effects on the temperature at $t_{i}$. However, due to the rapid thermal conduction, heat generated during the interval $t_{i-j-1}$ to $t_{i-j}$ has already spread throughout the entire WD, so the influence of energy lost before $t_{i-j}$ extends over the entire WD. 

\begin{figure*}[ht]
	\centering
	\vspace{-0.15in}
	\begin{minipage}{1\linewidth}	
		\subfigure{
			\label{fig:1}
			\includegraphics[width=0.49\linewidth,height=2.5in]{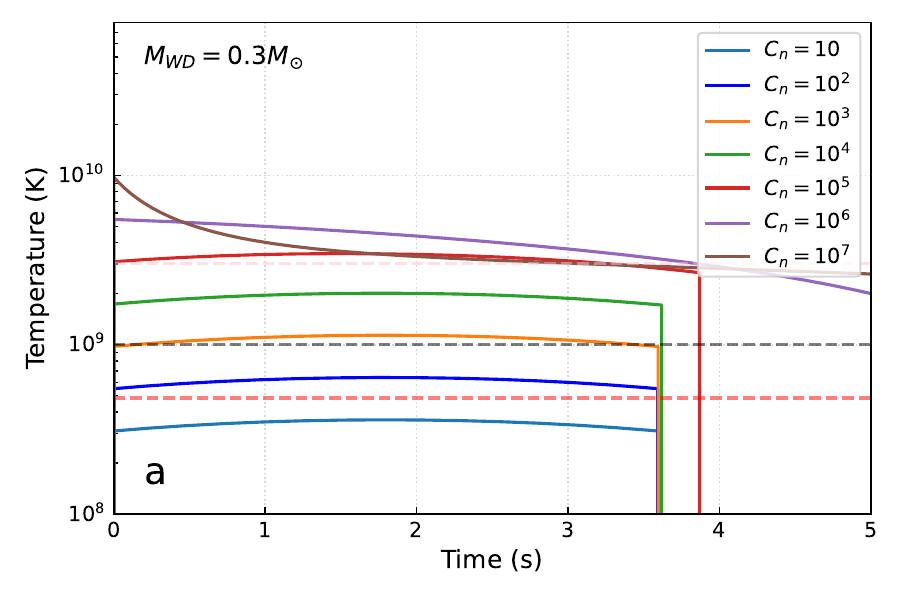}	
		}\noindent
		\subfigure{
			\label{fig:2}
			\includegraphics[width=0.49\linewidth,height=2.5in]{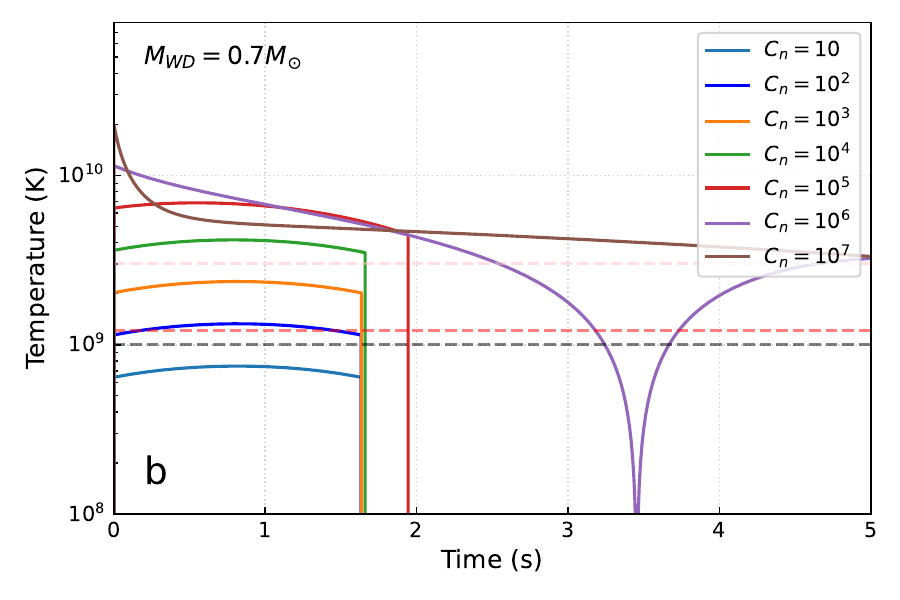}
		}
	\end{minipage}
	\vskip -0.8cm 
	\begin{minipage}{1\linewidth }
		\subfigure{
			\label{fig:3}
			\includegraphics[width=0.49\linewidth,height=2.5in]{{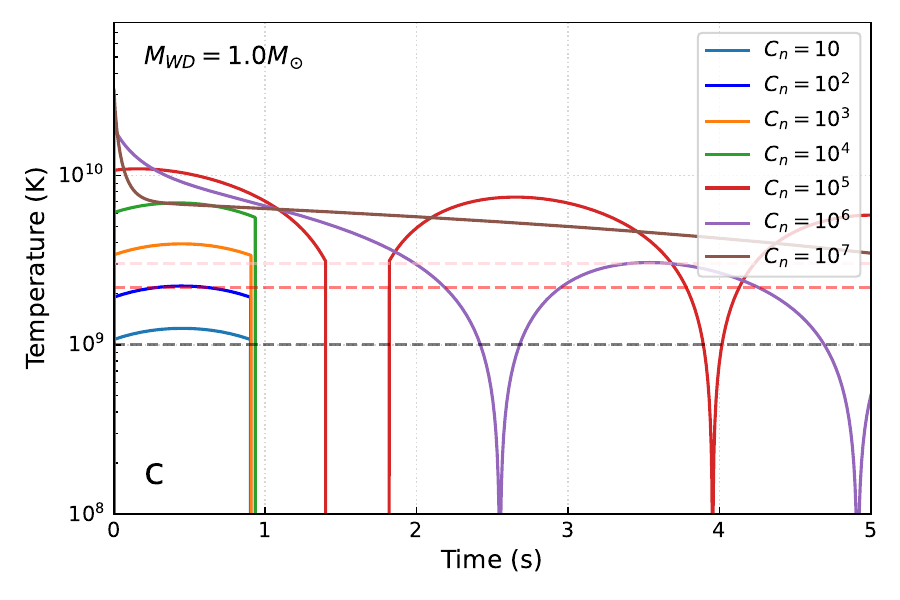}
			}
		}\noindent
		\subfigure{
			\label{fig:4}
			\includegraphics[width=0.49\linewidth,height=2.5in]{{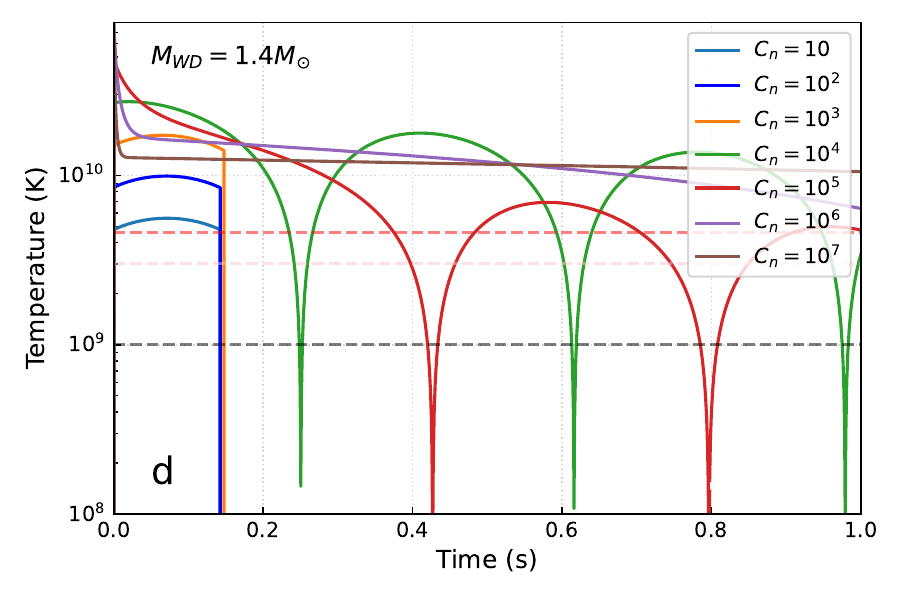}
			}
		}
	\end{minipage}
	\vspace{-0.18in}	
	\caption{The temperature profile of the reaction region. The sharp drop in the curve corresponds to the NS traversing the WD. The black dashed line denotes a temperature of $1\times 10^{9} \rm K$, which is the ignition temperature of carbon\citep{Chen2014}, while the pink dashed line represents $3\times 10^{9} \rm K$, corresponding to the critical burning temperature of oxygen \citep{Wu2018}. The red dashed line denotes the temperature at which the thermal pressure equals the electron degeneracy pressure.}
	\vspace{0.1in}		
	\label{fig:Temp}
\end{figure*}

Figure \ref{fig:Temp} shows the evolution curves of the temperature of the material at the interaction region as the NS traverses the interior of the WD under different parameter conditions. The sudden drop in the temperature curve indicates the moment when the NS exits the WD. In Figure \ref{fig:Temp}c, the red solid line first drops to zero and then reappears with a new temperature curve. This behavior is consistent with the discussion in Section 2.1, where under certain parameter conditions, the NS, after exiting the WD, can be pulled back by gravity and re-enter the WD. This occurs because our calculations assume idealized conditions for the NS-WD interaction.

\section{The result of the collision between an NS and different types of WD}

In discussing the collision between an NS and a WD, it is important to consider that the outcome will vary depending on the type of WD involved, due to differences in composition and structure. WDs can be classified into three types based on their primary constituents: helium WDs (He-WDs), carbon-oxygen WDs (CO-WDs), and oxygen-neon WDs (ONe-WDs).

\subsection{The NS collide with a CO-WD}
Due to the large population of CO-WDs, they are the most likely type of WD to collide with an NS in the universe. CO-WDs have a broader mass range compared to the other two types, ranging from $\sim$0.3 $M_\odot$ to $\sim$1.4 $M_\odot$, with the peak mass estimated to be in the range of 0.6-0.7$M_{\odot}$, according to the statistical results of \citet{Kepler2007}. Figure \ref{fig:Disp} shows the evolution curves of the relative distance between the NS and the WD under different parameter conditions. As the smaller the mass of the WD, the larger its radius, consequently, the relative velocity of the NS during its collision with the WD is reduced, resulting in a longer transit time for the NS through the interior of the WD. When the effective drag coefficient $C_{\rm n}$ is large, the drag force between the two objects may sufficiently decelerate the NS, potentially preventing it from penetrating the WD. Figure \ref{fig:coll} presents the maximum separation between the NS and the WD within 10 seconds after their contact under different parameter conditions. The red line indicates the NS was just able to pierce through the WD. It can be seen that for WDs with larger masses, the $C_{\rm n}$ value required for the NS to pierce through the WD decreases. However, even for WDs with masses approaching the Chandrasekhar limit, a $C_{\rm n}\sim 10^{4}$ is still necessary to decelerate the NS sufficiently to prevent it from penetrating the WD. This implies that in a head-on collision between an NS and a WD, even without considering whether the collision could trigger a thermonuclear explosion of the WD, forming a TZlO system as proposed by \citet{Pas2011a} requires at least $C_{\rm n}\gtrsim 10^{4}$. If the WD mass is taken as $M_{\rm WD}\sim 1.0~M_{\odot}$, then $C_{\rm n}\gtrsim 10^{5}$ is required.

The collision process between the NS and the WD, as well as the subsequent transit of the NS through the WD, is inevitably accompanied by intense energy conversion. Assuming negligible deformation of the NS during the interaction, the system's total kinetic energy and potential energy are converted into each other, with a fraction transferred into the turbulent and internal energy of the local WD material interacting with the NS. Here, we assume that 10$\%$ of the converted energy contributes to the internal energy. For a degenerate star like a WD, with high thermal conductivity, the localized internal energy will rapidly dissipate throughout the entire star. Assuming the internal energy is carried away entirely by electrons moving at relativistic speeds, we can calculate the temperature of the WD material at the reaction region at each moment. The resulting temperature evolution is shown in Figure \ref{fig:Temp}. 

A larger effective drag coefficient of the WD implies stronger interaction between the NS and the WD, resulting in a greater fraction of energy being converted into internal energy and thereby leading to a higher temperature at the collision region. As shown in Figure \ref{fig:Temp}, the initial temperature of the collision region increases with $C_{\rm n}$. However, when $C_{\rm n}$ becomes sufficiently large to significantly decelerate the NS, the interaction region's temperature rapidly decreases. Eventually, the NS undergoes quasi-periodic oscillations near the center of the WD, and its temperature evolution curve also exhibits quasi-periodic behavior.

The black dashed line in Figure \ref{fig:Temp} indicates $T = 1.0\times 10^{9} \rm K$, which corresponds to the threshold temperature for unstable carbon ignition in CO-WDs \citep{Chen2014}. The red dashed line indicates the temperature at which the thermal pressure equals the electron degeneracy pressure. As shown in Figure \ref{fig:Temp}a, for low-mass WDs the carbon ignition temperature is higher than the temperature for lifting of degeneracy. For more massive WDs, however, the lifting of degeneracy temperature exceeds the carbon ignition temperature. Consequently, only within a limited parameter range can the collision-induced temperatures reach the carbon ignition threshold while the material remains degenerate, thereby allowing unstable nuclear burning. For example, when $M_{\rm WD} = 0.7M_\odot$, this condition is satisfied only within a very narrow interval of $C_{\rm n}\lesssim 10^{2}$ . When $M_{\rm WD} = 1.0\,M_\odot$, the condition holds for $10 < C_{\rm n} < 10^2$. For WDs approaching the Chandrasekhar mass, the collision temperatures under our chosen parameters already exceed the lifting of degeneracy temperature. However, for sufficiently large $C_{\rm n}$ (e.g.,$C_{\rm n} = 10^5$), the strong drag force rapidly reduces the relative velocity between the WD and the NS, weakening the interaction and causing the temperature in the collision region to decrease. At some point, the temperature may then fall into the range between the carbon ignition and degeneracy thresholds, which appears to permit thermonuclear runaway. In cases where the collision temperature simultaneously exceeds both the carbon ignition and degeneracy temperatures, it remains uncertain whether a thermonuclear explosion would occur. In our calculations, the spatial scale of the heated region is comparable to the NS radius. Although the resulting high temperature may locally lift degeneracy, it is not yet clear whether nuclear burning would release sufficient energy to further lift degeneracy in the surrounding material, or instead ignite the still-degenerate layers and trigger a thermonuclear explosion. Taken together, our results indicate that for moderately massive WDs, there exists a parameter space in which can satisfy the ignition conditions for thermonuclear explosions.

When considering the conditions for thermonuclear explosions, the formation of a TZlO differs from the results discussed earlier. Taking the case of $M_{\rm WD} = 1.4\,M_{\odot}$ as an example, Figure \ref{fig:Temp}d shows that when $C_{\rm n} > 10^4$, the collision between the NS and the WD can effectively prevent the NS from penetrating through the WD, and the initial post-collision temperature is high enough to lift degeneracy. However, as the interaction weakens, the temperature in the collision region gradually decreases, and within a certain time interval it falls into the range that satisfies the ignition conditions for a thermonuclear explosion. In this case, the WD would be outburst, preventing the formation of a TZlO. 

For low-mass CO-WDs, as shown in Figure \ref{fig:Temp}a, the electron degeneracy-lifting temperature is lower than the carbon ignition temperature. As a result, the conditions required for a thermonuclear explosion cannot be satisfied. If $C_{\rm n}>5\times 10^{5}$, the NS is unable to escape from the WD and will eventually remain embedded within it, leading to the formation of a TZlO.

In addition to this scenario, alternative pathways for TZlO formation may exist in cases where the NS is able to emerge from the WD. During its passage through the WD, the NS disrupts the WD material in the interaction region and, owing to its strong gravitational field, drags along part of the WD’s matter. After the NS exits the WD, this material may remain gravitationally bound to it, giving rise to a TZlO. However, in this case the envelope mass of the TZlO would be smaller than in the former scenario, and its subsequent evolution would likely differ accordingly. This mechanism for the formation of TZlO is equally applicable to the cases of He-WDs and ONe-WDs.

\subsection{The NS collide with a He-WD}

He-WDs are typically of low mass, $\sim$0.2$\,M_\odot$. If formed through single-star evolution, their progenitors would need to have masses below 0.8$\,M_\odot$, which results in lifetimes that can exceed the current age of the universe. Therefore, He-WDs formed through single-star evolution are unlikely to exist in the current universe and are typically found in binary or multiple-star systems. \citet{Madej2004} identified a few candidates for He-WDs with masses around 0.2$\,M_\odot$ in a sample of over a thousand WDs from the Sloan Digital Sky Survey. Subsequently, \citet{Maxted2014} discovered 17 additional He-WD candidates from the Wide Angle Search for Planets data. These findings suggest that He-WDs are extremely rare and generally reside in multi-star systems.

Although He-WDs constitute a negligible fraction of the overall WD population, the discovery of several candidates makes it worthwhile to briefly discuss the collision scenario with an NS. Assuming $M_{\rm WD} = 0.3~M_{\odot}$, the $C_{\rm n}\gtrsim 5\times 10^{5}$ is required to slow down the NS sufficiently, as seen in Figure \ref{fig:Disp}a. The temperature evolution curve in the interaction region following the collision is shown in Figure \ref{fig:Temp}a. Due to the larger radius, lower density, and smaller relative velocity between the He-WD and the NS compared to higher-mass WDs, the interaction is relatively weaker, resulting in a lower temperature in the collision region.

However, the ignition temperature for helium ($T_{\rm ign} \sim10^{8}~\rm K$, \citet{Nomoto1977}) is significantly lower than that required for carbon ignition. For a He-WD with a central density of $1.0\times 10^{5} \rm ~g~cm^{-3}$, this threshold temperature can be readily achieved under the parameter conditions considered here. As shown in Figure \ref{fig:Temp}a, when $C_{\rm n}<30$, the collision-induced temperature exceeds the helium ignition threshold while the white dwarf material remains degenerate, thereby triggering a thermonuclear explosion of the WD. In contrast, when $C_{\rm n}>30$, the collision produces temperatures higher than the electron degeneracy-lifting threshold, and thus a global thermonuclear explosion of the star may not occur. However, it should be noted that our current understanding of the physical composition of WDs remains limited, and even less is known about He-WDs. Whether an NS collision can trigger the nuclear burning of a He-WD requires further verification through future numerical simulations.

\subsection{The NS collide with an ONe-WD}

ONe-WDs are typically the products of more massive progenitor stars, with masses generally exceeding $1.0~M_{\odot}$, and they represent a small fraction of the WD population. Due to their primary composition of oxygen and neon, it is traditionally believed that when they reach the Chandrasekhar limit, they do not undergo a thermonuclear explosion like CO-WDs. Instead, they undergo electron-capture reactions, ultimately collapsing into an NS, a process known as accretion induced collapse (AIC). However, some studies suggest that the energy released from the electron-capture process in neon could be sufficient to ignite oxygen, triggering an oxygen deflagration that could potentially lead to a partial or complete explosion of the ONe-WD \citep{Nomoto1984, Isern1991, Jones2016}.

In this paper, we explore the collision process between NSs and WDs. Due to the rarity of ONe-WDs, their collision with NSs is less probable than that of CO-WDs. Considering the larger mass of ONe-WDs, particularly when they approach the Chandrasekhar limit (as shown in Figure \ref{fig:Disp}d), if the parameter $C_{\rm n}$ is less than $\sim 10^{4}$, the NS will be able to penetrate the WD. However, when $C_{\rm n}> 10^{4}$, the NS will be unable to penetrate and will remain within the WD. Nevertheless, as shown in Figure \ref{fig:Temp}d, under all parameter conditions, the temperature in the interaction region during the collision can reach approximately $3\times 10^{9}~\rm K$. At this temperature, the oxygen can be ignited \citep{Wu2018}. However, in cases where the temperature exceeds the oxygen ignition threshold, it also surpasses the electron degeneracy-lifting temperature. Only when the NS is unable to emerge from the WD—for instance, when $10^4 \lesssim C_{\rm n}\lesssim 10^5$—does the weakening of the interaction cause the collision temperature to evolve through the range between the oxygen ignition and degeneracy-lifting thresholds. Within this interval, despite not reaching the conditions for electron-capture reactions, the collision between the ONe-WD and the NS will still result in an oxygen deflagration wave. However, because the conditions for electron-capture reactions are not met, this will ultimately lead to a thermonuclear explosion of the WD, rather than an AIC process. Thus, the collision of a massive ONe-WD with an NS does not trigger AIC but may instead lead to a special type of SN Ia.

For collisions between less massive ONe-WDs and NSs, as shown in Figure \ref{fig:Temp}c, the degeneracy-lifting temperature of the WD is lower than the oxygen ignition temperature. Consequently, igniting oxygen burning inevitably requires the corresponding material to become non-degenerate, thereby preventing a thermonuclear explosion of the WD. For $C_{\rm n} \lesssim 10^5$, the NS eventually emerges from the WD after the collision, whereas for larger values of $C_{\rm n}$, the NS ultimately remains trapped within the WD, leading to the formation of a TZlO.

\section{Summary and Discussions}

This study investigates the special case of a head-on collision between an NS and a WD. In this process, since the WD is a low-density object compared to the NS, we treat the WD as a fluid and the NS as a rigid body. After the collision, the NS drills into the WD, and the larger the mass and density of the WD, the greater the viscous resistance exerted on the NS. However, since WDs are degenerate matter objects, their drag coefficient $C_{\rm d}$ is not well known, and the strong gravitational field of the NS causes its collision cross-section with the WD to be larger than the geometric cross-section of NS. Therefore, we use an effective drag coefficient ($C_{\rm n}$) as a free parameter for our analysis.

As shown in Figure \ref{fig:coll}, under certain parameter conditions, the NS cannot penetrate the WD, while under other conditions, the NS can directly pass through the WD. We analyzed the temperature variations in the interaction region of the NS-WD collision under different parameters. The results indicate that as the mass of the WD and $C_{\rm n}$ increase, the temperature generated by the collision also increases. Different types of WDs, due to their varying compositions, have different ignition temperatures, which in turn leads to varying outcomes in their collisions.

\begin{itemize}
\item For He-WDs: A thermonuclear explosion of a He-WD can only be triggered when the collision-induced temperature exceeds the helium ignition threshold but remains below the electron degeneracy-lifting temperature, which requires that $C_{\rm n}$ not be too large. Given the relatively low mass of He-WDs, the resulting supernova is comparatively dim and features shorter rise and decline times. When $C_{\rm n}$ takes larger values, the collision temperature exceeds the electron degeneracy-lifting threshold, leading only to localized stable nuclear burning of the white dwarf material rather than a thermonuclear explosion. However, if $C_{\rm n}$ is sufficiently large that the NS cannot penetrate through the WD, then as the interaction weakens, the temperature in the collision region may fall into the range required to trigger a thermonuclear explosion.

\item For CO-WDs: as CO-WDs constitute the majority of the WD population, their collision probability with NS is higher, and their mass distribution is broader. We separately computed the collision scenarios between NSs and WDs of varying masses. In cases where the WD mass and $C_{\rm n}$ are larger, the collision is more intense. Figure \ref{fig:coll} shows the parameter conditions under which the NS can exit the WD during a collision. Figure \ref{fig:Temp} displays the temperature curves from the collision and compares them with the ignition temperature for carbon and the electron degeneracy-lifting temperature. For less intense collisions, the temperature may not reach the threshold for carbon ignition, and the WD is unable to prevent the NS from penetrating through its interior. In cases where the temperature in the interaction region consistently exceeds the electron degeneracy-lifting temperature. In these cases, the WD may have two possible fates: one is that similar to WD–main-sequence star collisions, the strong shock wave generated by the collision completely disrupts the WD \citep{Van2024}; the other is that the shock wave is insufficient to destroy the WD. However, regardless of the outcome, if the carbon is not ignited, no significant observational signatures are expected.

For violent collisions, where the high temperatures produced are sufficient to ignite the carbon, and this portion of the material remains degenerate, a Type Ia supernova may occur. Moreover, because most WDs are sub-Chandrasekhar limit, the thermonuclear explosion produces a smaller amount of $Ni$, and the ejecta mass is also lower compared to normal Type Ia supernovae. The collision result in significant deformation of the WD, with the ignition occurring off-center. This off-center ignition is anticipated to lead to pronounced asymmetry in the supernova. In addition, the energy injection of the collision causes the supernova to explode at a higher speed. Consequently, the NS collision-induced Type Ia supernovae are expected to be lower luminosities, faster light curve evolution, higher velocities, and significant optical polarization. Additionally, since the velocity of the supernova ejecta exceeds that of the NS, whether or not the NS could be decelerated sufficiently, a high-velocity NS is likely to remain within the supernova remnant.

\item For ONe-WDs: For low-mass ONe-WDs (e.g., $1.0~M_\odot$), the electron degeneracy-lifting temperature is lower than the oxygen ignition temperature. Thus, when the collision produces temperatures high enough to ignite oxygen burning, the material is already in a non-degenerate state, preventing a thermonuclear explosion of the WD. For more massive WDs, however, the temperature required for degeneracy lifting becomes higher and eventually exceeds the oxygen ignition threshold. In such cases, the collision can heat the interaction region to temperatures above the oxygen ignition condition while the material remains degenerate, thereby making a thermonuclear explosion of the WD possible.
\end{itemize}

Our results indicate that: (i) for He-WDs, since their electron degeneracy-lifting temperature is always higher than the helium ignition temperature, collisions with NSs either trigger a thermonuclear explosion of the WD or allow the NS to penetrate through without being captured, thereby preventing the formation of a TZlO; (ii) for CO-WDs, in the case of massive WDs, the electron degeneracy-lifting temperature exceeds the carbon ignition temperature. Thus, if the NS is to be retained within the WD, the weakening of the interaction inevitably leads the collision region temperature into the regime required for a thermonuclear explosion, making the formation of a TZlO impossible. For lower-mass CO-WDs, however, the electron degeneracy-lifting temperature is below the carbon ignition temperature, and if $C_{\rm n}$ is sufficiently large, the NS can be efficiently decelerated, resulting in the formation of a TZlO \footnote{Here, we are considering the simplified case where the condition for thermonuclear explosion due to degeneracy lifting does not apply. In reality, when the collision temperature exceeds both the electron degeneracy-lifting and nuclear ignition thresholds, whether or not a thermonuclear explosion occurs depends on the competition between the nuclear burning rate and the degeneracy-lifting rate: if nuclear burning proceeds faster, the WD is disrupted, whereas if degeneracy is lifted more rapidly, no thermonuclear explosion takes place.}; (iii) for ONe-WDs, the situation is similar to the CO-WD case, for low-mass WDs, where the electron degeneracy-lifting temperature is below the oxygen ignition temperature, a sufficiently large $C_{\rm n}$ allows the formation of a TZlO. In contrast, for more massive ONe-WDs where the degeneracy-lifting temperature exceeds the oxygen ignition threshold, collisions with NSs cannot produce a TZlO regardless of the value of $C_{\rm n}$.

The collision rate between WDs and NSs is a challenging quantity to estimate and is undoubtedly extremely low. The most likely systems in which such events might occur are triple or higher-order multiple systems, with the highest probability in regions of elevated stellar density, such as in clusters and galactic centers. A similar process is the head-on collision of two WDs. For example, \citet{Katz2012} calculated the probability that the Kozai-Lidov mechanism in triple systems can lead to WD–WD collisions and provided a very optimistic estimate, suggesting that the WD-WD collision rate could be comparable to the observed rate of Type Ia supernovae. In contrast, \citet{Toonen2018} presented a more conservative estimate, indicating that WD-WD collisions might account for only about 0.1$\%$ of the total Type Ia supernova rate. Given that NSs are much less abundant than WDs, NS–WD systems are even rarer than WD–WD systems and the collision cross-section between an NS and a WD is also smaller than that for two WDs. Thus, even under the most optimistic assumptions, the collision rate between NSs and WDs is extremely low, occurring only as a highly sporadic event in regions of exceptionally high stellar density. 

We have presented a toy model to calculate the dynamic process for collisions between NS and WD. However, should such events occur, they would involve an extremely complex set of physical processes. For instance, as the NS approaches the WD, significant deformation of the WD may be induced; the high relative velocity during the collision would generate shock waves, leading to turbulence in the local WD material; due to the inherent structural heterogeneity of WDs, variations in density distribution across different regions can result in different dynamical effects and ignition conditions; both NSs and WDs are systems with strong gravitational fields, making it necessary to consider general relativistic effects during the collision; and since both types of objects exhibit strong magnetic fields, a magnetohydrodynamic description of the interaction would be more accurate. These issues are all critical and warrant further investigation through detailed numerical simulations in the future.

\begin{acknowledgements}
We thank Y.-F. Huang and G.-M. Wang for helpful discussions about this paper. This work is supported by the National Natural Science Foundation of China (Projects 12373040,12021003), the National SKA Program of China (2022SKA0130100), and the Fundamental Research Funds for the Central Universities.

\end{acknowledgements}


\end{document}